# Delving into Transition to the Semantic Web

**Ioan Despi,**
University of New England, Armidale, Australia
**Lucian Luca,**
"Tibiscus" University, Timisoara, Romania

**REZUMAT**. Tehnologiile semantice reprezintă noua provocare pentru felul în care construim și exploatăm sistemele informatice. Ele sunt instrumente pentru a reprezenta semnificații, asociații, teorii, separat de date și de cod. Scopul lor este de a crea, descoperi, reprezenta, organiza, prelucra, gestiona, raționa, prezenta, partaja și utilizae semnificații și cunoștințe în vederea îndeplinirii unor scopuri de afaceri, personale sau sociale.

## 1 Introduction

Heraclitus' philosophy is expressed by his famous "panta rei" (πάντά ρέι) - *everything is changing* – and so is today's World Wide Web, which is undergoing a fast transition towards and gives way to tomorrow's Semantic Web (SW). Yesterday's World Wide Web represented information in such a way that humans could process it easily to infer new facts and to feed their senses by reading in a natural language (usually English), watching at graphics and listening to audio files. The creator of the Web, Tim Berners-Lee, described the World Wide Web as a combination of Hypertext and Internet, where documents can link to other documents, disregarding their location, thus producing webs of documents without central control or central repository. The next and natural change was to make dynamic content available on behalf of user interaction, and to incorporate new information into the created documents. The shift from static to dynamic Web was made possible by means of *web applications*, programs or scripts used to generate the dynamic content, to handle interactions with it and to deliver the result with a web server. Web applications are part of the World Wide

Web, because they can be accessed by means of the WWW. For humans, to combine data on the web was easily, even if different terminologies were used on different places. But the web is an open and ever-growing environment in which the amount of information and the number of available services make performing these tasks quite challenging for human users. The vast amount of information contained in the Web today is beyond human abilities to process it and because it was produced and targeted to human consumption, it is not suitable for automatic semantic interpretation. The Semantic Web together with personal software agents and web services provide the today solution to this challenge by allowing content to be annotated with machine understandable semantic descriptions.

The Semantic Web is not a new web but rather a logical evolution of the previous (current?) one, it is about "web evolution, not web revolution" (Tim Berners-Lee). It is a combination of knowledge representation using semantic networks and Internet. As machines do not have a comparable human ability to combine data or to use unstructured or partial information, not to mention to extract meaning from an image or an audio file, the Semantic Web should add machine readable and understandable information to the Web. Hence, the main distinction between the classical and the Semantic Web lies in the representation of information – syntax level versus semantic level. Going further, instead of having isolated repositories of knowledge, each using their own concepts and semantics, the Semantic Web will share and link the common concepts and their semantics in a web of data and metadata without a central control or a central repository. One can use the term *semantic web application* to name the pattern of using a program or a script to access, process and publish semantic web data and to handle interactions with data by using a web server to deliver the resulting information. Therefore, semantic web applications can be seen as a generalization or superset of web applications. The goal of the Semantic Web is to create an universal medium for information exchange by giving meaning to the content of documents on the Web. The vision is to make Web resources more understandable by machines, so SW extends the WWW through the use of standards, markup languages, logic inferences and related processing tools to produce, process and deploy machine-understandable annotations that will describe the content and functions of the Web resources. The Web would become a place where data can be shared and processed by automated tools as well as by people in a transparent manner. The rest of the paper is organized as follows: in section 2 we present the concepts of semantics and ontology, section 3 deals with Semantic Web standards, Web Services are described in section 4; the main criticisms are exposed in section 5 and we conclude in section 6.

## 2 Semantics

Searching through billions of documents on the Web for the information what we want is returning to us just data that must be interpreted by humans to be of any use. Semantic technologies (standards and methodologies) help humans to discover more explicit meaning from the data at their disposal. Semantics has always been important in computer science. What has changed with Semantic Web is the scale, as we can't design a system and define all its semantics locally. Semantic technologies are functional capabilities that enable both users and machines to use meanings and knowledge separately from data and program code.

Semantics is the science of meaning, answering questions such as "what do you mean when you say mouse?", "how do you associate meaning with symbols?", "what if a word has multiple meanings?" and so on. To explain the relationship between concepts, symbols and objects, Odgen and Richards [OR23] used the semiotic triangle invented by C.S. Pierce. Its vertices are Concepts, Symbols (signifier) and Objects (referent, signified). Symbols can only link to Objects through Concepts, in other words a symbolic representation of an object can never refer directly to objects (no real edge in the triangle). Actually, it can be represented by a linear composition *symbol* ↔ *concept* ↔ *referent*. From the point of view of SW it is interesting to mention that communication involves exchanging symbols that describe common attributes of objects, so we are interested in finding appropriate one-to-one mappings of attributes that describe a common symbol. The main problems here are standards and usage (both from vague to precise), as well as the fact that the same symbol may have different meanings in different places. Also, semantic mappings are relative in time and between groups of users. The Semantic Web is an enhanced Web where HTML pages are augmented with data understandable by the machines. A first step was represented by the use of XML as an alternative encoding to HTML such that the machines still can translate it in human format. XML is a representation language in which one can declare and use simple data structures, but not complex knowledge. New improvements such as XML Schema address this issue but still are not completely adequate to represent and reason about the knowledge involved in the Semantic Web. After an incredible amount of research and development, the World Wide Web Consortium (W3C) endorsed RDF, RDFS and OWL as World Wide Web standards in the area of semantics. Their adoption sent a strong message to vendors and users that from now on all products and technologies in the field of Semantic Web in order to be compatible should obey these standards.

The term Ontology migrated into AI from philosophy (theory of existence) and can be defined as "a specification of a conceptualization" [Gr93], where a conceptualization is an abstract view of the world (domain of discourse) and a specification is a systematic account of it. It contains objects, concepts and relationships between them and it represents a shared description of a domain of discourse. Hence it is used for communication purposes and this makes it different from a data model. Gruber chose to represent an ontology as a set of definitions over a formal vocabulary so an ontological commitment is an agreement to use the same vocabulary to make assertions and ask queries between different agents. An ontology deals with explicit descriptions of concepts (called *classes*) in a domain of discourse, properties (*slots*) of each concept (its features and attributes), as well as restrictions on properties (*facets*). An ontology together with some instances of classes represents a *knowledge base*. Inheritance is allowed too, so a class can have *subclasses* that represent concepts that are more specific than the *superclass*.

## 3  Semantic Web Standards

They are based on a vision of Tim Berners-Lee (for the most famous figure, see slide 10 of http://www.w3.org/2000/Talks/1206-xml2k-tbl) in which the basics are represented by Unicode and URI [BLFM05]. They are fundamental because the former provides an unified way to write texts in any alphabet known, while the latter provides the ability to uniquely identify resources and relations between them. The Extensible Markup Language [BPSa04] on top of these is also a fundamental component (together with namespaces and restricted with XML Schema) that provides the fundamental syntax for representing relationships and meaning. The Resource Description Framework [MM04] is a family of standards using URI and XML to unambiguously make statements about resources by means of relationships and meaning, thus creating a data model for objects. The data model is represented in XML syntax. Anything that can be identified by URI could be described and be in relationship with anything else. The metadata model makes statements about resources in the form of a Subject (resource), a Predicate (property) and an Object (value), that is by means of a *(subject, predicate, object*) tuple, called a *triple* in RDF terminology. URIs in XML tags and property names can be abbreviated by using XML-namespaces. A value (object) can be a resource or a literal, while a literal is a string that cannot be the subject of a statement, and therefore has no properties. A property (predicate) relates subjects to objects and it is a resource because it may or may not have properties of its own. A Statement is also a resource as it can have properties (e.g., who, when was created the statement). RDF also enables us to create Vocabularies (terms) in which some statements should be used. The terms would indicate that a given RDF describes a specific class of resource with use of specific properties. Usually, such vocabulary is provided by RDF Schema [BG04]. RDFS extends RDF to define domains and ranges, class and property hierarchies, comments and labels. It still does not allow the description of constraints. DAML (DAML + OIL) extends RDFS by adding description logics to it. It successor, Web Ontology Language [Ma04] layers on top of RDF adding more vocabulary for describing properties and classes, e.g., relationships between classes (disjointness, cardinality), equality, features of properties (symmetry), and enumerated classes. The upper layers of the Semantic Web are still under construction so they are not very well standardized. Description Logics are used in knowledge representation to express concepts and their hierarchies. They are a subset of FOL which is function-free and does not allow explicit variables. Regarding digital signatures, Semantic Web uses asymmetrical ciphers with public and private keys and one-way hashing functions. The last two levels in the overall picture – trust and proof – are still to be devised.

## 4 Web Services

Web Services standards are designed to make possible communications and exchange of data via the Internet machine to machine, without human intervention. They are independent to any operating system, programming language or transport protocol. Once again, one can represent their architecture by means of a triangle, whose vertices are Service Provider, Service Registry and Service Consumer. A Service Provider implements a service and describes it in Web Service Description Language. This description (together with some contact information) is submitted to a Service Registry (by that Service Provider) and registered there as an entry such that any Service Consumer interested in the usage of a particular service queries the registry to obtain the WSD. Then the consumer can generate the appropriate message to communicate with the provider over a standard protocol. Web Services technology relies on a pile of XML based protocols, e.g., Web Service Description Language (WSDL), http://www.w3.org/2002/ws/desc, Simple Object Access Protocol (SOAP), http://www.w3.org/TR/soap, and Universal Discovery, Description and Integration (UDDI), http://www.uddi.org. The communications are based on passing XML messages in a common document format. To operate a Web Service, a user sends an XML message containing a request for the Web Service to perform some operations and, in response, the Web Service sends back another XML message containing the results of the operation.

WDSL documents are the basic bricks of this architecture: each one is based on a XML schema which describes the XML instance and it defines services as collections of network endpoints (called ports). The root element of a WDSL is identified by definitions and it can contain up to seven different types of sub-elements (types, message, operation, port type, binding, port, service). Anyway, a complete WSDL definition should contain all of the information necessary to invoke a Web Service.

Developers that want to allow others to access their services should make WSDL definitions available [Sk03]. UDDI is a registry based on XML for Web Services to list themselves on the Internet such that the consumers can find them. In this way the companies can find each other (name, product, location) and make their systems interoperable for e-commerce. The project was started by Microsoft, IBM, and Ariba and now includes more than 130 companies. An alternative to the public nature of registry is to make a private agreement between a service consumer and a provider by sharing a specific WDSL file.

The SOAP protocol is used for communications between programs on different operating systems on the Web by using the Hypertext Transfer Protocol (HTTP) and its Extensible Markup Language (XML) as the mechanisms for information exchange. More precisely, SOAP specifies how to encode an HTTP header and an XML file so that programs in different computers can call each other and pass information. One can say that HTTP is SOAP's RPC-style transport, and XML is its encoding scheme. In general, the SOAP protocol is responsible to deliver XML messages from one place to another on the Web. A SOAP message is a one-way transmission between two SOAP nodes, from a sender to a receiver, and contains three main parts: an envelope, a header and a body. The exchange of SOAP messages is done by means of "underlying" protocols and this specification is called SOAP binding. SOAP is a central element of Microsoft's .NET architecture for future Internet application development. Web Services are tools that can be used in many different ways, most prominent being RPC, SOA and REST.

The Service Oriented Client (SOC) is a person expected to interact with the above services. Service Oriented Clients complement the capabilities of SOA and are focused directly on the services they need instead of the structure of objects that implement these services. SOAP is intensively used for serialization/deserialization, i.e., the process of transforming a runtime object into a stream of data and vice-versa.

## 5 Criticisms

The first concern found in the literature [Ma06] is that it is unrealistic to expect busy businesses to create enough metadata to make the Semantic Web work. Semantic Web applications require a complicated infrastructure, in which terms are ordered based on their conceptual relationships and then fit into the resulting schema. To create, maintain or adapt these schemas is not a trivial task. The solution to this problem may simply be better tools for creating metadata. The way in which we build metadata can strongly influence the future of Semantic Web. A well-known essay [Do01] raises important concerns on the different flavors of metadata becoming widely adopted and used within the Semantic Web.

The second concern relates to the ability of ontologies (designed to translate between different data descriptions) to realistically help computers understand the concepts, even the basic human ones (usually poly-semantics, e.g., dwelling).

The third main concern is about the name Web Services. Actually, web services do not have much to do with the web besides using the internet as a transport medium and therefore relying on protocols such as HTTP, FTP or SMTP. The criticisms from the step-brother REST community envisage the improper usage of URIs and of the state-less architecture of the web [FT02]. Going further, the Web Services do not obey one of the main principles of the web (which states that publication of information should be based on a global and persistent URI), being stateful conversations based on the hidden content of messages [MNS04].

Clay Shirky believes that "the Semantic Web is a machine for creating syllogisms" [Sh03]. The Semantic Web is made up of assertions and it provides means of putting them on the Web, such that users can combine them to discover things that are true but not specified directly. He is critic to the role of deductive logic in every day living and to the Web too, blaming A. C. Doyle and his Sherlock Holmes hero for this misconception. Some of his ideas can be summarized as "meta-data is not a panacea" because it is untrustworthy, "ontology is not a requirement" – look at RSS autodiscovery, "worse is better" philosophy, as it is better to start with a minimal creation and grow it as needed (see also [Ga91]).

Dan Zambonini [Za06] identifies seven flaws, starting with the unequal creation of (meta) data ("all of that data being created by the life sciences, and other similar groups, is behind closed doors. It isn't available to the public, and doesn't hook into any web"), then noticing that "a technology is as good as developers thing it is" and comparing the number of books published about Semantic Web, Ajax, RDF, etc. The point is that "complex systems must be built from successively simpler systems" and not by burning phases, and that "a new solution should stop an obvious pain", which is not the case with the actual web. Creating metadata and classifications is difficult ("people are not perfect") but "you don't need an ontology of everything". But it would help to have a central ontology. He concludes that "philanthropy isn't commercially viable" so every user should have a serious reason to start exposing their data in a compatible format. Hence, the need for that killer application.

Tim Berners-Lee answers to some critics of the Semantic Web in an interview [Up05]. He thinks that many critics did not understand the philosophy of how Semantic Web works. Regarding a the most vehiculated idea that "the Semantic Web doesn't do anything for me I can't do with XML" he argues that this comes from "someone who is very used to programming things in XML, and never has tried to integrate things across large expanses of an organization, at short notice, with no further programming". Another source of confusion is the fact that people treat it as a big XML document tree so that they can use XML tools on it, when in fact it is a web, not a tree. His decision to use a web and not a semantic tree is based on the fact that a tree just doesn't scale (where is the root?) and only webs can be merged together in arbitrary ways. Tim agreed with criticisms of the RDF/XML syntax that it isn't very easy to read. And, by the way, he sees that killer app to be the Intranet for a company.

**Conclusions**

There is a lot missing from the current Web and the gap can be filled in part by using a logic-based framework. However, for any such project to be successful, compatibility with syntax-oriented HTML and XML systems must be kept. The future Web will undoubtedly be a compromise between the specifications of the researchers as W3C group and the real life demands of the practitioners like Oracle, Microsoft and other major actors. There is more to say about technical aspects of Semantic Web than we thronged here. Forthcoming Semantic Web standards, such as SPARQL query language and Gleaning Resource Descriptions from Dialects of Languages (GRDDL) mechanism for adding semantic annotations to XHTML pages will simplify the work of embedding semantics in other web applications, while funds from DARPA, NSF and IST will boost new large projects. Small enterprises will continue to add metadata by using RDF and OWL, so the entire picture is quite optimistic.